\documentstyle[prl,preprint,aps]{revtex}
\tighten
\begin{document}
\draft
\title{R-Process Freezeout, Nuclear Deformation, and the Rare-Earth Element 
Peak}
\author{Rebecca Surman$^{\rm a}$, Jonathan Engel$^{\rm a}$, Jonathan R. 
Bennett$^{\rm a,b}$, and Bradley S. Meyer$^{\rm c}$}
\address{$^{\rm a}$Department of Physics and Astronomy, University of North 
Carolina, Chapel Hill, NC 27599}
\address{$^{\rm b}$Bartol Research Institute, University of Delaware, Newark, 
DE 19716}
\address{$^{\rm c}$Department of Physics and Astronomy, Clemson University, 
Clemson SC 29634-1911}
\date{\today}
\maketitle
\begin{abstract}
We use network calculations of r-process nucleosynthesis to explore the 
origin of the peak in the solar r-process abundance distribution near nuclear 
mass number $A\approx 160$.  The peak is due to a subtle interplay of nuclear 
deformation and beta decay, and forms not in the steady phase of the 
r-process, but only just prior to freezeout, as the free neutrons rapidly 
disappear.  Its existence should therefore help constrain the conditions 
under which the r-process occurs and freezes out. 

\end{abstract} 


The r-process is responsible for synthesizing roughly half the heavy nuclei
in the solar system (see Refs.\ \cite{Cow,Mey94} for a review).  It is widely
believed to occur somewhere in core-collapse supernovae, at a time when the
density of free neutrons is so high that neutron capture by nuclei occurs
much more rapidly than nuclear $\beta$ decay.  Under these conditions
equilibrium between neutron capture and photodisintegration (called
$(n,\gamma)-(\gamma,n)$ equilibrium) establishes itself so that very
neutron-rich isotopes of each element are populated.  The nuclei eventually
$\beta^-$ decay, turning one of their neutrons into a proton, then resume
capturing neutrons until equilibrium is reached again.  This ``steady" phase
of the r-process continues as long as the free neutrons remain abundant.
When the neutrons begin to disappear, $(n,\gamma)-(\gamma,n)$ equilibrium
becomes more difficult to maintain and $\beta$ decays play a larger role in
determining the most abundant isotopes of a given element.  Eventually the
neutron-capture and photodisintegration reactions ``freeze out'', and the
nuclei simply $\beta$-decay back to the stability line.

Large peaks in the solar r-process abundance distribution at nuclear mass
numbers $A\approx 80$, 130, and 195 are apparently due to closed neutron
shells.  Closed-shell nuclei made during the r-process have strongly bound
neutrons and long $\beta$-decay lifetimes, causing their abundance to build
up.  The source of a smaller peak at $A\approx 160$, the region of rare-earth
elements (REEs), is far less clear.  The authors of Ref.\ \cite{Bur}
speculated the cause to be deformation of neutron-rich REE nuclei.  Beyond
the closed shell at neutron-number $N=82$, in this scenario, the stabilizing
effect of deformation allows many more neutrons to be captured.  When $A$
reaches about 160, however, the nuclei can deform no further, and so add
neutrons with difficulty, often $\beta$-decaying instead.  The nuclei
populated just past the deformation maximum therefore ought to be closer to
the valley of stability than their predecessors and consequently should have
somewhat longer $\beta$-decay lifetimes, leading to a moderate build up as
the r-process proceeds.  The calculations of \cite{See} lent this hypothesis
some support, but produced a much broader abundance hump than was observed
and was unable to reproduce other r-process features.

An alternative explanation, first proposed in Ref.\ \cite{Cam} and elaborated
in Ref.\ \cite{Sch}, is that the REE peak is produced by mass-asymmetric
fission of very heavy r-process nuclei.  The fission fragments form a
double-peaked abundance distribution in nuclear mass number, and the heavier
products are supposed to fill in the r-process abundance curve at the
location of the REE peak.  By now, however, this idea is not compelling.  As
neatly pointed out in Ref.\ \cite{Mar}, the r-process should terminate by
$\beta$-delayed fission before $A\approx 160$ fission fragments can produced,
and the small odd-even effect in the $A\approx 150-170$ region would be
erased by fission.  The source of the REE peak has therefore never been
satisfactorily explained.

Despite the lack of an explanation, recent r-process
simulations\cite{Mey92,Woo} in the high-entropy neutrino-driven bubble near
the surface of the remnant neutron star have produced a nice REE peak.  The
simulations used calculated nuclear properties (far from stability) in which
the effects of deformation had been included self-consistently.
Significantly, however, the simulations did not allow nuclear fission.  
Moreover, they were among the first to follow the r-process all the way 
through freezeout.  All this suggests that nuclear deformation, not fission, 
is responsible for the REE peak, but that the dynamics leading to freezeout 
rather than the mechanism of Ref.\ \cite{Bur} cause it to form.  In this 
paper we confirm and explicate this idea.

Our conclusions are based on an analysis of the development of the r-process
``path", which we define as follows:  At any given time, for each element 
with proton number $Z$, there is an isotope with maximum abundance; 
the collection of all such isotopes defines the path at that time.  Before 
freezeout, when $\beta$-decay rates are much less than neutron-capture or 
disintegration rates and the system is in $(n,\gamma)-(\gamma,n)$ 
equilibrium, the path's location follows from the requirement that the free 
energy be stationary under the transfer of a neutron from a nucleus to the 
free-neutron bath.  In the approximation that the nuclei and free neutrons in 
the r-process are ideal gases this condition, which is equivalent to detailed 
balance, implies that for every $Z$ the isotope with maximum abundance has a 
neutron separation energy given by
\begin{equation} 
S_n (Z, N_{max}) = -kT\ln\left \{ {\rho N_A Y_n
\over 2} \left ({2\pi\hbar^2 \over m_n k T}\right )^{3/2} \right \}~,
\label{eq:pathloc} 
\end{equation} 
where $N_{max}$ is the neutron number of the isotope, $T$ is the temperature, 
$\rho$ is the mass density, $N_A$ is
Avagadro's number, $Y_n$ is the abundance of free neutrons per nucleon, and
$m_n$ is the neutron's mass.  Eq.\ (\ref{eq:pathloc}) implies that in 
equilibrium the path
always lies along a contour of constant separation energy.

The high-entropy r-process calculations of Refs.\ \cite{Mey92,Woo}, which
showed a REE peak, were performed within detailed simulations of supernova
explosions.  Here we simplify matters by ignoring supernova fluid dynamics,
instead parameterizing the dependence of temperature on time (with $\rho
\propto T^3$) to roughly match the results of the more complicated
calculations.  We then use a computer code developed at Clemson University to
solve the differential equations (described in Ref.\ \cite{Cow}) that
determine the time-development of nucleosynthesis.  The inputs, besides the
temperature and density, are the initial mass fractions of neutrons and 
preexisting ``seed"
nuclei, and calculated neutron capture rates\cite{Cow}, neutron separation
energies\cite{Mol}, and $\beta$-decay rates\cite{Mol}.  The seed nuclei
quickly come into $(n,\gamma)-(\gamma,n)$ equilibrium and then undergo the
usual $\beta$-decay and neutron-capture sequence.  We run the simulations
through freezeout until only stable nuclei remain.  The treatment of
freezeout is fully dynamical in that neutron capture and photodisintegration
continue to compete with $\beta$ decay throughout the entire process, even
when equilibrium no longer obtains.  In the more detailed supernova
simulations, the high-entropy r-process occurs over several seconds, and many
r-process components combine to give the final abundance distribution (see,
e.g., figure 15 of Ref.\ \cite{Woo}).  The calculations presented here treat
just the components with the highest initial neutron/seed nucleus ratios,
because they are responsible for the REE peak.

A sample set of predicted r-process abundances appear in figure 1a alongside
the measured solar-system abundances.  In this run the initial seed nucleus
was $^{70}$Fe and the initial value of $R$, the ratio of the abundance of
free nucleons to that of nuclei, was 57, implying that a seed
captured on average 57 neutrons.  The very poor agreement just above the peak
at $N \approx 82$ ($A \approx 130$) is unexplained but apparently plagues all
such simulations.  It may be due to a deficiency in the nuclear mass
extrapolation\cite{Mey93} or the neglect of a second component with
slightly different temperature and neutron density.  In any event, we wish to
draw attention here to the presence of a REE peak at approximately the
correct location and with the correct width.  No sign of this peak exists
during the steady phase of the r-process (figure 1b).  It forms only after
$R$ falls below about 1, when steady $\beta$ flow is destroyed and the path
begins to move towards stability.  The peak appears under a wide range of
initial temperatures, densities, and neutron/seed ratios provided only that
freezeout from equilibrium is prompted by the capture of nearly all free
neutrons rather than a sudden drop in overall density and temperature, as is
often assumed.  The primary reason is that, surprisingly, the r-process stays
in approximate $(n,\gamma)-(\gamma,n)$ equilibrium even after steady flow
fails.  Only when $R$ has fallen by many orders of magnitude does $\beta$
decay completely dominate the other reactions and destroy the remnants of
$(n,\gamma)-(\gamma,n)$ equilibrium.  As a result, the path continues for
some time to lie roughly along contours of constant neutron separation energy
even as $\beta$ decay moves it moves towards stability.

Figure 2 illustrates this phenomenon; it shows the path between $N=82$ and
126 at three times during a 0.25-second interval just after the steady phase
ends.  During this period, as indicated by the insert in figure 2, $R$ (or
$Y_n$) drops dramatically (the diamonds mark the three times at which the
path is plotted).  The dark squares in the large figure are the paths as
defined above, with the upper set corresponding to the later time.  The open
diamonds indicate the ``equilibrium paths'', i.e.\ those that would obtain if
the system were in true $(n,\gamma)-(\gamma,n)$ equilibrium according to Eq.\
(\ref{eq:pathloc}).  Contours of constant neutron separation energy for the
even $N$ nuclei are overlaid.  The actual and equilibrium paths indeed differ
very little well after the end of the steady phase.  Eventually, as the
figure shows, a ``kink'' in the separation energies at $N\approx 104$ is
intercepted and imposes itself on the path.

Large kinks in the path are the underlying cause of the abundance peaks at
the neutron closed shells mentioned above.  Because the neutrons in all these
nuclei are strongly bound the path turns sharply up towards stability when a
closed shell is reached, producing a concentration of populated isotopes
close together in $A$ with relatively long $\beta$-decay lifetimes.  The
early explanation of the REE peak in Ref.\ \cite{Bur} is a variation on the
same theme.  In our simulations, however, the REE peak does not form in
exactly this way, even though it is clearly associated with the $N \approx
104$ kink, which in turn is clearly due\cite{Mol} to the deformation maximum
postulated in Ref.\ \cite{Bur}.  The nuclei at the top of the kink, and thus
closer to stability, do have moderately slower $\beta$-decay rates than those
along the path immediately below, but neither this fact nor the proximity of
the kink nuclei to one another along the path account entirely for the
pronounced REE peak.  Another mechanism, connected with the motion of the
path as it traverses the kink, is also at work.

In the vicinity of the kink the average separation energy that determines the
equilibrium path grows at a rate governed by the beta decay of an
``average-lifetime'' nucleus, which is typically in the kink.  Below the
kink, as just mentioned and illustrated in the center of figure 3, the nuclei
along the path are farther from stability and therefore decay faster than
average, before the average separation energy has changed enough to move the
equilibrium path in their vicinity.  In an attempt to return to the path and
stay in equilibrium, these nuclei then capture neutrons (which are rapidly
dwindling in number), increasing their value of $A$.  The nuclei above the
kink, by contrast, decay more slowly than average, and in general will not do
so before the equilibrium path at their location has moved.  When it does
move, these nuclei, whose thermodynamic impetus is also to remain in
equilibrium at the average separation energy, photodisintegrate so that their
mass number $A$ is lowered.  The net result, shown with arrows in the insert
in figure 3, is a funneling of nuclei into the kink region as the path moves
toward stability.

To confirm these ideas, we ran a simulation with our own simplified and
easily varied nuclear properties.  We obtained binding energies from a simple
semi-empirical mass formula (\cite{sim}) and $\beta$-decay lifetimes by
fitting those of Ref.\ \cite{Mol} with a function of the form $T_{1/2}=a
Q^{\alpha}$, where $Q$ is the difference in binding energies between the
parent and daughter and the best fit was obtained with $a=250$ and
$\alpha=-3.80$.  We assumed neutron capture rates, the details of which are
irrelevant, to have an exponential dependence on separation energy,
reflecting their dependence on the level density in the compound nucleus
formed by capture.  We started the simulation at the end of the steady phase
of the r-process, with abundances along the equilibrium path between $N=82$
and $N=126$ taken to be identically normalized Gaussians of width 1.05
centered at $N_{max}$ for each $Z$.  These conditions produced a flat final
abundance curve, shown in figure 4a alongside the (appropriately scaled)
solar abundances.  When we introduced a kink into the separation energies at
$N=104$, adjusting the capture and $\beta$-decay rates to reflect the new
bindings, a REE peak matching that in the solar-abundance curve formed
(figure 4b).  But when we modified the $\beta$-decay rates to be constant
along curves of constant separation energy, destroying the
$\beta$-decay-induced dynamic described above, the peak shrank by a factor of
about 2 relative to the surrounding abundance level (figure 4c).  This
smaller peak therefore represents the effects of the concentration of points
along the path and the slight increase in $\beta$-decay lifetimes at the top
of the kink.  These factors alone are clearly not enough to fully produce the
peak.

The simplified set of freezeout-related calculations just described
summarizes the necessary and sufficient conditions for the formation of a REE
peak with the correct size.  We conclude that this peak is indeed due to a
deformation maximum in the region, but that in order for it to form the
maximum must be traversed after the steady phase of the r-process ends but
before the system completely freezes out of $(n,\gamma) - (\gamma,n)$
equilibrium.

Nuclear physics has often contributed to our understanding of the r-process,
and vice versa.  Through the study of certain $\beta$-decay rates, for
example, the authors of Ref.\ \cite{Kra93} provided experimental support for
the idea that the r-process distribution contains many components.  On the
other hand, observed r-process abundances near $A \approx 120$ apparently say
something significant about the strength of the closed $N=82$ shell near the
neutron-drip line\cite{Chen}.  These important discoveries have all involved
the steady phase of the r-process (but see Ref.\ \cite{Haxton}).  By contrast
the REE peak is associated with freezeout, and is therefore sensitive to
conditions at later times.  The work reported here allows us to conclude, for
instance, that freezeout is prompted by the exhaustion of free neutrons
rather than a rapid drop in overall density and/or temperature.  This fact
may constrain models in which the expansion of the r-process region is so
fast that freezeout occurs before nearly all the neutrons have been captured.
The existence of the peak also confirms the predicted deformation of
neutron-rich rare-earth nuclei, and its fine structure may constrain nuclear
models.  Finally, the effect on the REE peak of neutrinos emitted from the
cooling supernova remnant has yet to be examined.  The delicate interplay of
nuclear deformation, neutron capture, photodisintegration, and $\beta$ decay
that forms the REE peak as the r-process freezes out may thus have more
to tell us about the conditions under which heavy-element nucleosynthesis
occurs.

The authors are grateful to D.H.\ Hartmann for initiating this collaboration
and to D.D.\ Clayton for valuable discussions.  This work was supported in
part by the U.S.\ Department of Energy under grant DE-FG05-94ER40827 and by
NASA under grant NAGW-3480.  We thank the Institute for Nuclear Theory at the
University of Washington, where some of this work was carried out.

\begin{figure}
\caption{Calculated (line) and measured (crosses) solar r-process abundances.  
Part a) shows the final calculated abundances and part b) the calculated 
abundances before freezeout, while the r-process is in the steady phase.}
\label{f:1}
\end{figure}

\begin{figure}
\caption{The r-process path between $N=82$ and 126 at three times 
during a 0.25-second interval early in the decay to stability. The shaded 
squares are the actual paths, the lightest corresponding to the latest time, 
and the diamonds the equilibrium paths defined at the same times
by Eq.\ (\protect{\ref{eq:pathloc}}).  The lines are contours of constant 
separation energy in MeV.  The insert shows the neutron abundance per 
nucleon over this interval, with the three times at which the path is 
depicted indicated by diamonds.}
\label{f:2}
\end{figure}

\begin{figure}
\caption{Contours of constant neutron separation energy in MeV (solid lines) 
and constant $\beta$-decay rates in $s^{-1}$ (dashed lines).  The insert is a 
schematic of two such contours, with the arrows depicting the flow of nuclei 
into the region containing the separation-energy kink.}
\label{f:3}
\end{figure}

\begin{figure} 
\caption{Abundances on a linear scale near $A=160$ in the
simplified model (see text).  Part a) is with smooth separation-energy
contours, part b) with a kink in the contours, and part c) with constant
$\beta$-decay rates along each kinked contour.  The crosses are the 
appropriately scaled solar abundances.} 
\label{f:4} 
\end{figure}

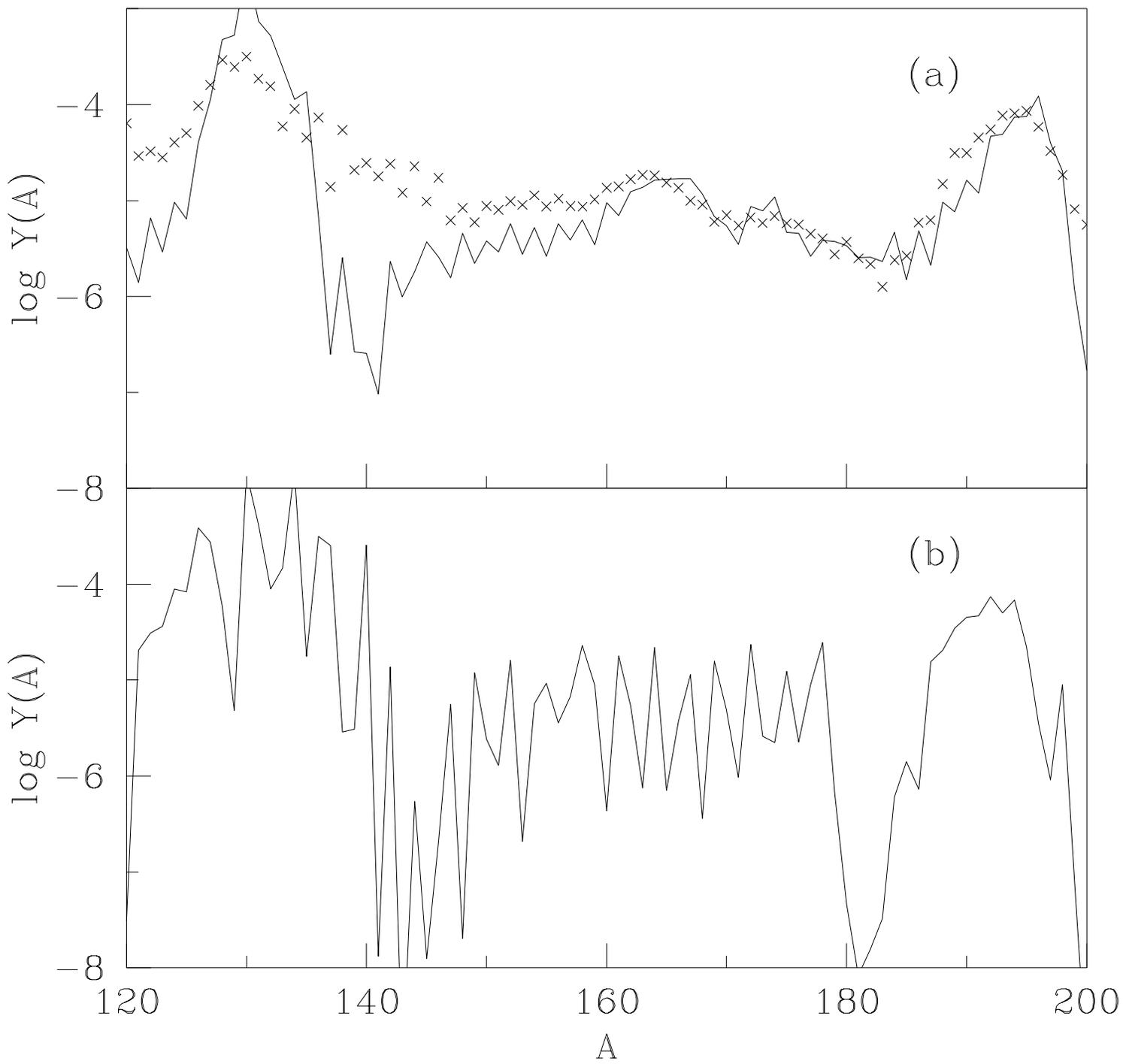
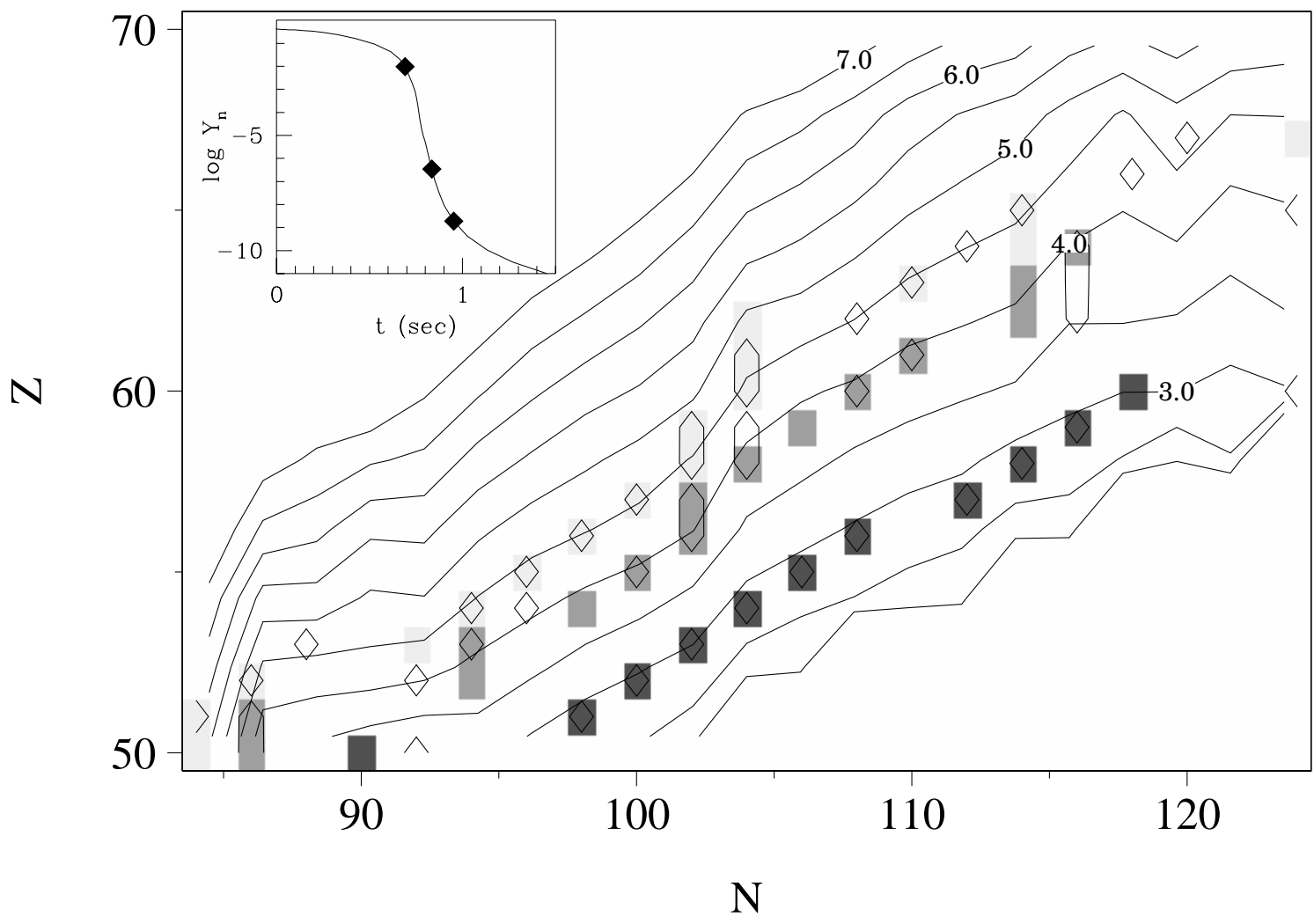
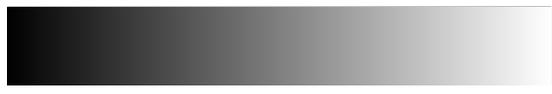
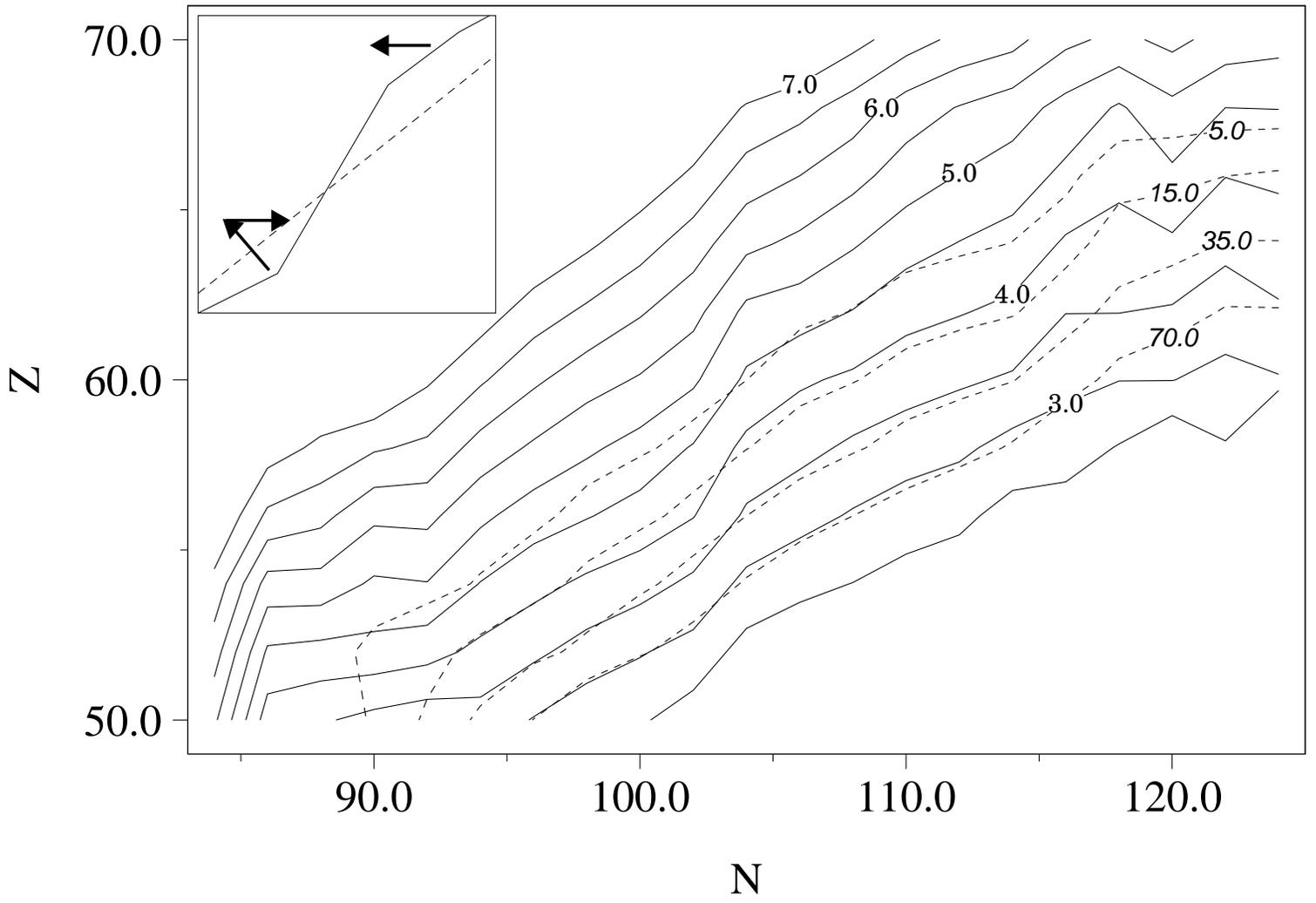
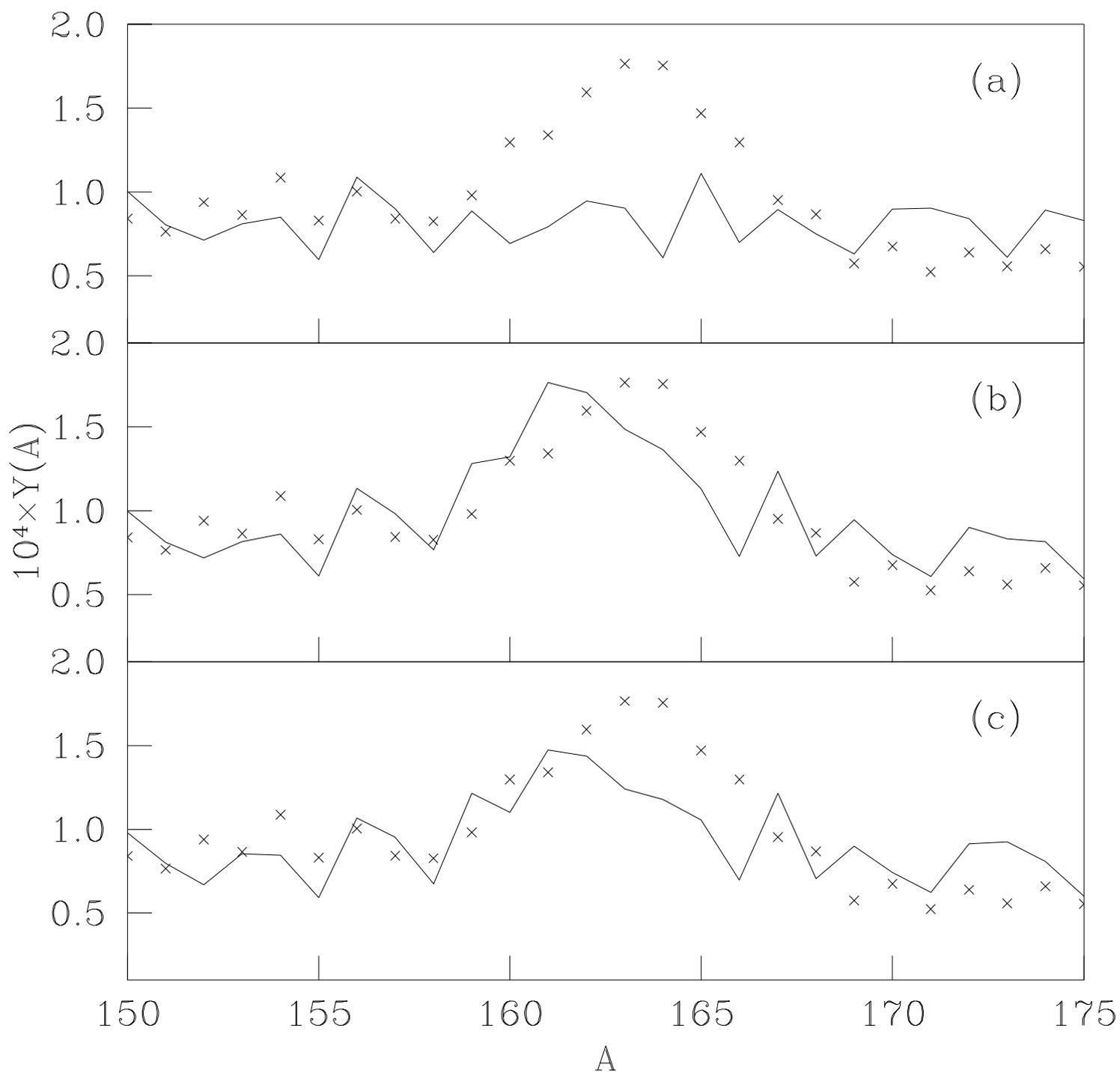
\end{document}